\documentclass[twocolumn,showpacs,preprintnumbers,amsmath,amssymb]{revtex4}

\usepackage{graphicx}
\usepackage{dcolumn}
\usepackage{bm}

\def \be{\begin{equation}}
\def \ee{\end{equation}}
\def \kup{k^\parallel_\uparrow}
\def \kdn{k^\parallel_\downarrow}

\def \ks{k^\parallel_{\uparrow,\downarrow}}

\begin{document}
\draft
\title{Current induced distortion of a magnetic domain wall}

\author{Xavier Waintal and Michel Viret}
\affiliation{CEA, Service de physique de l'\'etat condens\'e,\\
Centre d'\'etude de Saclay F-91191 Gif-sur-Yvette cedex, France\\}

\date{\today}

\begin{abstract}
We consider the spin torque induced by a current flowing
ballistically through a magnetic domain wall. In addition to a
global pressure in the direction of the electronic flow, the
torque has an internal structure of comparable magnitude due to
the precession of the electrons' spins at the "Larmor" frequency.
As a result, the profile of the domain wall is expected to get
distorted by the current and acquires a periodic sur-structure.
\end{abstract}
\maketitle

With the advent of "spintronics", which aims at using the spin of
charge carriers in devices, electronic transport in ferromagnets
is being revisited from a different viewpoint. The focus has been
for a long time on the effect of magnetism on transport properties
(e.g. Magneto Resistance, MR)~\cite{gmr}, but it is now realized
that the electronic current can be a tool to change the
magnetization direction. The relevant effect, known as spin torque
has attracted considerable interest recently in the context of
ferromagnetic--normal metal--ferromagnetic trilayers~\cite{slon}.
There, the first magnetic layer acts as a spin filter, and the
incident polarized electrons exert a torque on the second layer.
At the heart of this physical effect is the fact that spin
currents are not preserved when electrons cross a magnetic layer,
and as a result some angular momentum is transferred to the
magnetization. It was shown experimentally that for strong enough
current densities this mechanism can lead to magnetic
reversal~\cite{torques-experiment}. This demonstrates the
feasability of current controlled magnetic memory cells, but the
current needed for complete reversal might be too high for
industrial implementation. An alternative would be to use the
current to move a domain wall (DW) in between two stable
positions. The idea that a current can apply a force on a DW is
due to Berger \cite{berger1} in the seventies. It is the aim of
this letter to study in detail the spin torque exerted on a DW in
the presence of an electric current. Our main finding is that in
addition to a global pressure, the torque has a spatially
dependant component that will lead to a deformation of the DW in a
periodic sub-structure.

At the root of understanding the spin torque in a DW is the
question of what happens to the spin of a conducting electron when
going through a DW. Two extreme cases can be considered called
"interface" and "adiabatic". A very sharp domain wall (expected in
constrictions for example), can be treated as an interface on
which the electrons can be specularly
reflected~\cite{CabreraFalicov}. This leads to the giant magneto
resistance effect as a magnetic field will remove the DW and the
associated extra resistance. Because the electronic spin is
conserved during this process, no spin torque is exerted on the
DW. On the other hand, in a very long domain wall, the electron's
spin will adiabatically adapt itself to remain aligned with the
local magnetization under the effect of the "Larmor"
precession~\cite{viret}. There, no MR is
observed\cite{CabreraFalicov,berger2} but each electron going
through the DW will flip its spin and give a quantum $\hbar$ of
angular momentum to the wall inducing a global pressure.

Experimentally, the wall resistance (DWR) has been measured in
macroscopic systems where it is of the order of a few percent per
wall. This resistance results from a slight mistracking of the
conduction electrons' spins which mixes the majority and minority
channels within the wall~\cite{viret,levy,bauer1}. Numerous
measurements have been reported in the last few years finding both
positive \cite{viret,viret2} or negative \cite{ruediger} effects.
The spread in the experimental results probably reflects how
difficult it is to extract the DWR among other contributions in
series coming from domains (like the Anisotropic MR). Other
(ballistic) models were developed~\cite{tatara,bauer2} to explain
negative effects, but it was later recognized that proper band
structure calculations \cite{bauer3} are needed to get reliable
quantitative results. Nevertheless, recent clean experiments
\cite{viret2} in nanostructures of perpendicularly magnetized
materials (FePd) have demonstrated that the spin scattering models
developed in \cite{viret,levy} can account for the MR in domain
walls in 3d metals. Concerning torque effects, very recently
several groups have been able to push DWs with a
current~\cite{expcurrpress}.

In this letter, we proceed as follows. First, an heuristic
argument is given to explain the origin of the spatial structure
of the torque. Then, we introduce our (ballistic) model and point
out that some key features of the band structure must be taken
into account. Finally we calculate the spin torque and evaluate
the corresponding distortion of the DW profile.

\begin{figure}[h]
\vglue +0.45cm
\includegraphics[width=9cm]{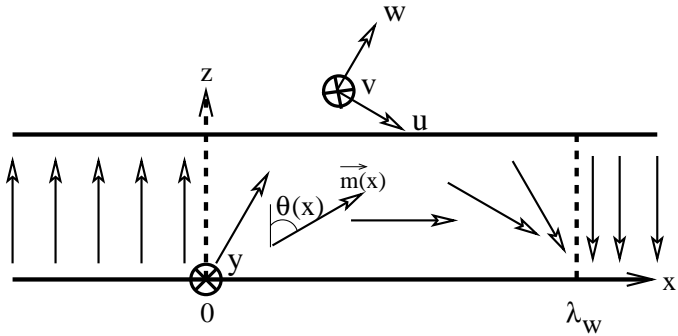}
\caption{\label{fig:system} Schematic of the domain wall of length
$\lambda_w$ and transverse dimension $d$. The arrows stand for the
magnetization direction $\hat m$ that makes an angle $\theta(x)$
with the $z$-axis. The same $x$, $y$ and $z$ axis are used for
both the space and spin basis for convenience. The rotating frame
$(u,v,w)=(d \hat m /d\theta,d \hat m /d\theta \times \hat m,\hat
m)$ follows the magnetization.}
\end{figure}

\begin{figure}[h]
\vglue +0.45cm
\includegraphics[width=8cm]{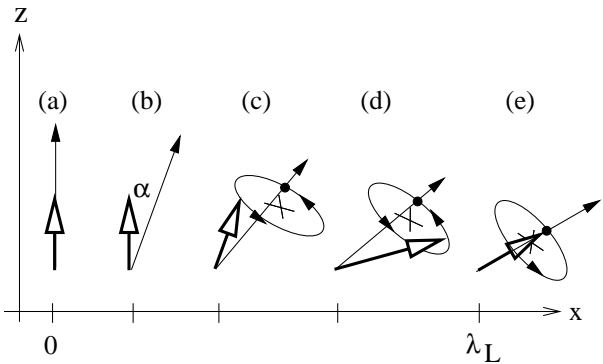}
\caption{\label{fig:schema} Cartoon of the effect of the Larmor
precession allowing the electron spin to follow the local
magnetization $\vec m$. The thin arrow represents $\vec m$ while
the thick one stands for the conducting electron spin.}
\end{figure}

{\it Heuristic argument}: Let us follow an electron going through
the (Neel) DW of size $\lambda_w$ sketched in Fig.\ref{fig:system}
(a Bloch wall is completely equivalent for MR and torque). The
process is described in Fig.\ref{fig:schema}: Before the wall, the
electron's spin is aligned with the local magnetization $\vec m$
(a). Then, when entering the DW, $\vec m$ begins to rotate, and a
small angle $\alpha$ starts to build up between the spin and the
local magnetization (b). As soon as $\alpha$ is not zero, the
electron's spin starts to precess around the direction of an
effective magnetization with a period equal to the "Larmor"
precession length $\lambda_L$ ((c) and (d)). At the end of every
period, the electron's spin is back onto the local magnetization
direction \cite{viret}. At $x=\lambda_L/2$, $\alpha$ reaches a maximum.
Over this distance, $\vec m$ has rotated by an angle
$\pi\lambda_L/2\lambda_w$ which thus gives an upper bound for
$\alpha$. Hence, $\eta=\lambda_L/\lambda_w$ is the parameter that
controls the crossover from the adiabatic limit ($\eta\ll 1$) to
the interface one ($\eta\gg 1$).

The (almost) adiabatic case is the most interesting regarding spin
torque because the incident spin up electron ends up with the
(almost) opposite spin after crossing the wall, losing an angular
momentum $\hbar$ in the process. In return, since the total
angular momentum is conserved, a torque of equal magnitude and
opposite sign is exerted on the magnetization within the wall.
This is the global pressure mentioned earlier. The spin precession
around the slowly rotating effective magnetization does not
conserve the angular momentum either, and, as detailed below, it
gives rise to the spatially dependant part of the torque inducing
a deformation with a period $\lambda_L$.


{\it Model.} Let us introduce a simple model for the conducting
electrons through the DW of Fig.\ref{fig:system}. The system is a
ferromagnetic wire running along $x$ and of typical transverse
dimension $d$. We approximate the wall by a linear rotation of the
local magnetization on a scale $\lambda_{w}$. The unit vector
$\hat m(x)$ lies inside the $xz$-plane and makes an angle
$\theta(x)=\pi x/\lambda_w$ with the $z$ axis inside the DW. The
Hamiltonian reads, \be \label{eq:H} H=-\frac{\hbar^2}{2m^*}\Delta
-\frac{J_{\rm exc}}{2}\ \hat m(x) \cdot \vec \sigma. \ee Here
$J_{\rm exc}$ is the exchange energy, $m^*$ the effective mass,
$\vec \sigma=(\sigma_x,\sigma_y,\sigma_z)$  the vector of pauli
matrices and the Fermi energy is noted $E_F$. The spin dependant
part of $H$ reads, \be \label{eq:theta} \hat m(x) \cdot \vec
\sigma= \left\{
\begin{array}{l}
\sigma_z,\ \ \ x<0\\
R_{\theta (x)}\sigma_z R_{-\theta (x)},\ \ \ 0<x<\lambda_w \\
-\sigma_z, \ \ \ \lambda_w<x\\
\end{array}
\right. \ee with the rotation matrix defined as $R_{\theta
(x)}=e^{-i\sigma_y \theta(x)/2}$. In the region outside the wall,
the eigenstates of $H$ are plane waves in both the transverse
$yz$-plane, with total momentum $k^\perp$, and in the
$x$-direction, with momentum $\kup$ ($\kdn$) respectively for
majority and minority electrons. In the transverse direction,
$k^\perp$ is quantized in units of $2\pi/d$ and for the
corresponding eigenstate to be propagating in the $x$-direction,
one needs $\ks=\sqrt{2m^*/\hbar^2 (E_F \pm J_{\rm exc}/2)
-(k^\perp)^2}>0$ which leads to $N_\uparrow$ ($N_\downarrow$)
propagating channels for the majority (minority) spin. Once these
propagating channels are defined, the natural way to calculate
physical quantities such as the current or the spin current is to
use the Landauer-Buttiker theory and fill up the different
eigenstates of the system. This has been widely used in the
literature and eventually leads to the Landauer formula for the
conductance $g=e^2/h\  {\rm Tr}\ tt^\dagger$ where $t$
is the transmission matrix of the system.
\begin{figure}[h]
\vglue +0.45cm
\includegraphics[width=6cm]{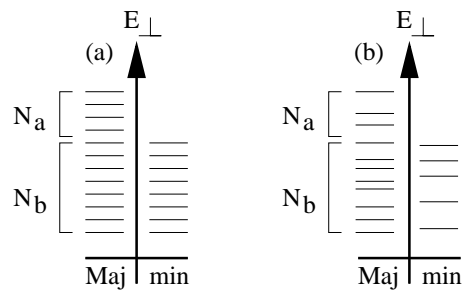}
\caption{\label{fig:band-struct} Cartoon of the position of  the
majority and minority channels as a function of the perpendicular
energy. The number of majority channels is $N_\uparrow=N_a+N_b$.
(a) simple picture where $N_b=N_\downarrow$. (b) ``realistic''
magnet where $N_b\ne N_\downarrow$.}
\end{figure}

In this simple picture, the $N_\downarrow$ down channels are
exactly matched by the first $N_b=N_\downarrow$ up channels with
equal perpendicular energy, and the polarization of the current is
entirely due to the remaining $N_a\equiv
N_\uparrow-N_b=N_\uparrow-N_\downarrow$ up channels that have
large perpendicular energy (see Fig.\ref{fig:band-struct} (a)).
Hence their longitudinal energy ($\sim\sqrt{J_{\rm exc}/m^*}$), is
low and the corresponding minority channels may not propagate. This stems
from the assumption that the up and down bands have exactly the
same shape, which is obviously erroneous. It was recently clearly
underlined by Mazin~\cite{mazim} that by oversimplifying the band
structure of magnetic metals, one introduces symmetries in the
system that ought not to be there. Indeed, when the polarization
of the current $P_I$ is taken equal to the polarization $P_N$
i.e., \be P_I\equiv\frac{I_\uparrow- I_\downarrow}{I_\uparrow
+I_\downarrow} =P_N \equiv\frac{N_a}{N_\uparrow +N_\downarrow}
   = \frac{J_{\rm exc}}{E_f}
\ee this artificial symmetry can lead to erroneous physics. For
instance, after an ``interface'' wall, all the channels with high
perpendicular energy are blocked (the $N_a=(N_\uparrow-
N_\downarrow)$ in Fig.\ref{fig:band-struct} (a)) and the
polarization of the current drops to almost zero. In a real system
however (sketched in Fig. \ref{fig:band-struct} (b)), the number
$N_a$ of blocked channels would not be equal to $(N_\uparrow-
N_\downarrow)$ and the system would retain some polarization
$P_I\approx(N_b- N_\downarrow)/(N_\uparrow- N_\downarrow)$. A
complete treatment of the problem would require ab-initio
calculations as described in~\cite{mazim}. Here, we break this
artificial symmetry by giving a different weight to the different
channels in the Landauer formula, i.e. allowing $P_I$ and $P_N$ to
be independent numbers. This can be viewed as replacing the
ferromagnet by a non-magnetic reservoir in series with a
normal-ferromagnetic interface that acts as a spin filter. It leads
to the following modified Landauer formula, 
\be \label{eq:cond-landauer} g=\frac{e^2}{h}
{\rm Tr}\ \left[ 1 + \frac{P_I-P_N}{1-P_N} \sigma_z \right]
tt^\dagger 
\ee 
where the $\sigma_z$ matrix applies in spin space and
only to those channels that can propagate for both majority and
minority spins. We emphasize that we introduce this modification to 
take into account the band structure of the magnet and do not question the
Landauer-Buttiker formalism itself. This correction is crucial here since, 
as will be seen later, by symmetry the spin torque vanishes when $P_N=P_I$.

For a given spinor wave function $\Psi(x)$, the spin current
flowing along the $x$-direction is defined as~\cite{wmbr}, 
\be
\vec J_{s}(x)=\frac{\hbar^2}{2 m^*}{\rm Im}\int dydz
\Psi^\dagger(x) \vec \sigma \frac{\partial}{\partial x} \Psi(x)
\ee 
This is not a conserved quantity, and the corresponding loss
of spin current is identified with the torque~\cite{wmbr}. The
global torque is then $\vec \tau_{\rm tot}=\vec J_{s}(x=0)-\vec
J_{s}(x=\lambda_w)$ while the local torque (per unit distance) is,
\be 
\vec \tau= -\frac{\partial}{\partial x}\vec J_{s}(x) 
\ee 
From the last equation, we derive for the torque per unit voltage:, 
\be
\label{eq:torque-landauer} \frac{\partial \vec \tau }{\partial V}=
-\frac{e}{4 \pi} \frac{\partial}{\partial x} {\rm Re \ Tr}\ \left[
1 + \frac{P_I-P_N}{1-P_N} \sigma_z \right] t(x) \vec \sigma
t^\dagger(x) 
\ee 
where the generalized transmission matrix $t(x)$
gives the amplitudes of the different modes inside the wall.


{\it Domain wall close to the adiabatic limit}. We now proceed
with the treatment of Eq.(\ref{eq:H}) and Eq.(\ref{eq:theta}) in
the limit of a long wall. The first step consists in writing the
schr\"odinger equation in the basis aligned with the local
magnetization. An eigenstate $\Psi(x)$ is then written as, \be
\Psi(x)=R_{\theta (x)} \Phi(x) \ee An effective equation is
obtained for $\Phi(x)$  (see~\cite{levy} for a similar treatment,
the exact solution of the model can be found using the ``spin
spiral'' state, see~\cite{bauer2}). The solution is then formally
expanded in series of $(1/\lambda_w)$  and we match $\Psi(x)$ and 
its derivative at $x=0$ and $x=\lambda_w$. 
In first order in $(1/Q_k\lambda_w)$ where
$Q_k=\sqrt{\kup\kdn}$, we
get for incoming majority electrons in the wall: \be
\Phi_\uparrow(x)=\frac{e^{i\kup x}}{\sqrt{\kup}} {\rm \ \ and \ \
} \Phi_\downarrow (x)= \frac{i\pi}{4Q_k\lambda_w\sqrt{\kdn}}\times
\ee
$$
\left( \frac{1}{P_k} [e^{i\kup x}-e^{i\kdn x}]+P_k [- e^{i\kup
x}+\xi e^{-i\kdn x}  ] \right)
$$
and similar solutions for incoming minority spins. The wave
functions are normalized to carry unit fluxes and only the
longitudinal part has been written. We have defined
$P_k=(\kup-\kdn)/(\kup+\kdn)$, and
$\xi=e^{i(\kup+\kdn)\lambda_w}$. We point out that the oscillatory
first term in the brackets is the expression of the Larmor
precession while the second one is a reflected wave. In the region 
$x<0$ the reflected wave takes the form, 
\be \Phi_\downarrow (x)=\frac{i\pi}{4}
\frac{P_k}{Q_k\lambda_w}(\xi-1) \frac{e^{-i\kdn x}}{\sqrt{\kdn}}
\ee

{\it Spin torque in a macroscopic DW.}
To proceed with the calculation of the conductance and torque, one
incorporates the expression for the wavefunctions into
Eq.(\ref{eq:cond-landauer}) and Eq.(\ref{eq:torque-landauer}) and
performs the average over transverse momentum. This average is
done in two steps, first over $\xi$, which is taken to be a random
phase and then over $k_\perp$. Also, the Fermi wave length
$\lambda_F$ and Larmor precession length $\lambda_L$ are defined
as,
 \be
 \lambda_F=2\pi\sqrt{\frac{\hbar^2}{2m^*E_F} },\ \
\lambda_L=\pi\sqrt{\frac{E_f}{J_{\rm exc}}}
\sqrt{\frac{\hbar^2}{2m^*J_{\rm exc}} } \ee

We get for the correction to the conductivity due to the presence
of the DW,$\Delta g/g=- P_N \lambda_F^2/(64\lambda_w^2)$. 
This correction is (too) small because in a ballistic model, the
calculated quantity is the reflected part of the wavefunction due
to the potential step. Diffusive models, on the other hand,
neglect this contribution and estimate the resistance due to spin
mixing between the up and down electrons. It turns out that in the
macroscopic case, the latter dominates. In the interface limit
however, the ballistic contribution becomes important. We
calculate the torque per current along the local frame $(u,v,w)$
from the wavefunctions expressions as,
$$
\frac{\partial \tau_u (x)}{\partial I} =
\frac{\hbar}{e}\frac{\pi}{2\lambda_w} \left[ P_I + (P_I-P_N) 
\cos\left(2\pi\frac{x}{\lambda_L}\right)\right],
$$
\be 
\label{eq:torque} \frac{\partial \tau_v (x)}{\partial I} =
-\frac{\hbar}{e}\frac{\pi}{2\lambda_w}(P_I-P_N)
\sin\left(2\pi\frac{x}{\lambda_L}\right). 
\ee 
Equation (\ref{eq:torque}) is the
central result of this letter. The torque consists of two terms.
The first one, proportional to $P_I$ pushes the wall in the
direction of the electrons and does not depend on $x$. The second
part of Eq.(\ref{eq:torque}) is much more interesting since it
leads to a deformation of the wall on the scale of $\lambda_L$.
Note that although its net contribution to the global torque is
small, its intensity is of the same magnitude as the first term. 
However, the $N_a$ channels where the
perpendicular energy is high (hence the down electrons cannot
propagate) do not contribute to the torque since no precession is
possible inside those channels. As a result, this component of the
torque vanishes when $P_I=P_N$, hence the necessity of taking into
account the fact that $P_I\neq P_N$ in real materials. We point out that
although our calculation is done for a ballistic model, the result should hold 
for realistic domain walls. Indeed, both the spin diffusion length
(about 50 nm for Ni) and the mean free path (a few nm) are larger
than $\lambda_L$ (of the order of 3 nm for Ni) and at the Larmor length
scale, the electrons can thus be considered ballistic. In a system where 
the mean free path would be smaller than $\lambda_L$ but still with a (rather)
large spin diffusion length, our conclusions would remain qualitatively 
correct, though a quantitative treatment might be needed.

In order to quantify the expected deformation of the wall, one
would have to solve self-consistently the torque equation together
with the reaction from $\hat m(x)$ linked to the "stiffness" of
the wall (using for instance the Landau-Lifshitz-Gilbert
equation). Here we simply evaluate the magnitude of the
deformation by comparing the energy given to the wall via the
torque to the total wall energy. Typical quantities for Ni are
$\lambda_w=100 nm$, $\lambda_f=0.2 nm$ , $\lambda_L=3nm$, and
$P_I-P_N\approx 0.5$. A current density $j=10^{10} A/m^{2}$ would
then leave in a Ni wall $10^{-5}J/m^{-2}$, which is one tenth of
the wall surface energy. Hence the distortion in angular gradient
should be of the order of several \%. This is a significant effect
considering that the chosen current density is one order of
magnitude below that used in tri-layer spin torque experiments.

{\it Constriction "interface" limit.} In a constriction (small
transverse direction $d$), the DW width is expected to scale with
$d$ \cite{PBruno,Labaye} and the system could be driven to the
interface limit. As explained earlier, the spin of the transferred
electrons would remain mostly un-flipped and the torque would
decrease as a result. Hence, while in this regime the MR gets
larger, the torque described in this paper would drop. In
addition, since in this limit the reflection is higher, another
contribution to the total pressure could come from the transfer of
momentum from the reflected electrons to the DW. Although it is
not clear to us whether this change of momentum would actually
push on the wall or on the atoms, its effect would be much smaller
than that in a long wall.

{\it Conclusion.} The torque generated by a current on a domain
wall in a ferromagnetic metal has been studied in unconstrained
DWs. It is composed of two contributions of comparable magnitude.
The first one is a global pressure resulting from the loss of
angular momentum of electrons crossing the wall. The second is due
to the precession of electrons' spins inside the wall which
generate a periodic torque. The first effect can be used to move
the DW with a current. Depending on pinning, the current density
necessary to dislodge the DW could be smaller than that necessary
to reverse a full magnetized layer in a spin valve. The effect can
then potentially be useful in spin electronic devices where a DW
switched between two stable positions could be used in the gate of
a three terminal device. Moreover, the periodic torque will
distort the wall's internal structure in a significant manner when
current densities of the order of $10^{10} A/m^2$ are driven through
it. This may help the depinning process and might also be able to
switch the wall between different types (Bloch and Neel for
example) in a similar manner to the predicted temperature effect
in a constriction \cite{Labaye}. The distortion may be measured by
several techniques including polarized neutron reflectivity (PNR)
and domain wall resonance. We also infer that a current flowing
parallel to the wall would produce a surstructure of the same
period, which may be easier to measure with PNR.

{\it Acknowledgment.} It is a pleasure to thank R. Jalabert and
D. Weinmann for interesting discussions.



\begin{references}

\bibitem{gmr}  A. Barthelemy, A. Fert, J.P. Contour, et al. J. Mag. Magn. Mat.
242: 68 (2002).

\bibitem{slon} J. Slonczewski, J. Magn. Magn. Mater. {\bf 159}, L1 (1996).

\bibitem{torques-experiment} E. B. Myers, D. C. Ralph, J. A. Katine, and
{\it al}, Science {\bf 285}, 867 (1999). J.  A. Katine, F. J.
Albert, R. A. Buhrman, and ,{\it al},  Phys. Rev. Lett. {\bf 84},
3129 (2000).

\bibitem{berger1} L. Berger, Phys. Lett. A {\bf 46A}, 3 (1973).

\bibitem{CabreraFalicov} \bibinfo{author}{G.~G. Cabrera} and \bibinfo{author}{L.~M. Falicov},
  \bibinfo{journal}{Phys. Stat. Sol.} \bibinfo{volume}{\textbf{61}},
  \bibinfo{pages}{539} (\bibinfo{date}{1974}).

\bibitem{viret} M. Viret, D.Vignoles, D. Cole and {\it al},
{\it Phys. Rev.} B  {\bf 53}, 8464 (1996).

\bibitem{berger2}
\bibinfo{author}{L.~Berger}, \bibinfo{journal}{{J. Appl. Phys.}} \bibinfo{volume}{\textbf{49}},
  \bibinfo{pages}{2156} (\bibinfo{date}{1978}).

\bibitem{levy} P.M. Levy and S. Zhang
{\it Phys. Rev. Lett.} {\bf 79}, 5110 (1997).

\bibitem{bauer1} A. Brataas, G. Tatara and G.E.W. Bauer
{\it Phys. Rev.} B  {\bf 60}, 3406 (1999).

\bibitem{ruediger}
\bibinfo{author}{U.~Ruediger}, \bibinfo{author}{J.~Yu},
  \bibinfo{author}{S. Zhang}, \bibinfo{author}{A.~D. Kent}, and
  \bibinfo{author}{S. S. P. Parkin}, \bibinfo{journal}{{Phys. Rev. Lett.}}
  \bibinfo{volume}{\textbf{80}}, \bibinfo{pages}{5639} (\bibinfo{date}{1998}).

\bibitem{tatara} G. Tatara and H. Fukuyama,
{\it Phys. Rev. Lett.}  {\bf 78}, 3773 (1997).
\bibitem{bauer2} R. P. van Gorkom, A. Brataas and G. E. W. Bauer 
{\it Phys. Rev. Lett.}  {\bf 83}, 4401 (1999).

\bibitem{bauer3} J. B. A. N. van Hoof, K. M. Schep, A. Brataas and {\it al},
{\it Phys. Rev.} B {\bf 59}, 138 (1999).

\bibitem{viret2} M.~Viret, Y.~Samson, P.~Warin, et al.
  {\it Phys. Rev. Lett.} {\bf 85}, 3962 (2000);
  R. Danneau, P. Warin, J.P. Attane, et
al. {\it Phys. Rev. Lett.} {\bf 88}, 157201 (2002).


\bibitem{expcurrpress} N. Vernier et al., D. Ravelosona et al., R. Cowburn et al., A. Fert et al. {\it private communications}.

\bibitem{mazim} I. I. Mazim, {\it Phys. Rev. Lett.} {\bf 83}, 1427 (1999).

\bibitem{wmbr}  X. Waintal, E. B. Myers, P. W. Brouwer, and
D. C. Ralph  {\it Phys. Rev.}  B {\bf 62}, 12317 (2000).
 X. Waintal and P. W. Brouwer
{\it  Phys. Rev.} B {\bf 65}, 054407 (2002).

\bibitem{PBruno}P. Bruno, Phys. Rev. Lett. 83, 2425 (1999).


\bibitem{weinmann} D. Weinmann, R. L. Stamps and R. A. Jalabert,
in proceedings of the `36$^{\rm th}$ Rencontres de Moriond, edited by
T. Martin, G. montambaux and J. Tr\^ an Thanh V\^ an.

\bibitem{Labaye}  Y. Labaye, L. Berger and J.M.D. Coey {\it J. Appl. Phys.}
 {\bf 91}, 5341 (2002).






\end{references}
\end{document}